# Effects of the Position Reversal of Friction Pairs on the Strength of Tribocharging and Tribodischarging

Na Li,[a] Liran Ma,[b] Xuefeng Xu*[a] and Jianbin Luo[b]

[a] *School of Technology, Beijing Forestry University, Beijing 100083, China*
[b] *State Key Laboratory of Tribology, Tsinghua University, Beijing 100084, China*

## Abstract

   The friction-induced charging (i.e., tribocharging) and the following discharging (referred here as tribodischarging) are always believed to have negative effects on the daily life and on the industrial production. Thus, how to inhibit the tribocharging and the tribodischarging has caused wide public concern. Because the discharge caused by the electrical breakdown of the ambient gas is generally accompanied with the generation of light, we investigated here the tribocharging and the tribodischarging by observing the light emitted during friction. We found that the position reversal of the friction pair has a dramatic impact on the intensity of the tribo-induced light. Experimental results show that an intense light is produced when a stationary $Al_2O_3$ disk is sliding on a rotating $SiO_2$ disk, but only a weak light is observed for the case of a stationary $SiO_2$ disk and a rotating $Al_2O_3$ disk. This means that the process of the tribocharging and the tribodischarging can be significantly influenced owing to the change in the relative position of the friction couple. The experimentally measured polarities of the tribo-induced charge on the friction surfaces further indicated that the strong discharging occurs when the rotating surface is negatively charged. The reason for the difference in the intensity of the tribocharging and tribodischarging can be attributed to the combined effects of the contact potential difference and the temperature gradient between the contacting surfaces on the charge transfer when friction. Finally, a simple, low cost, yet effective approach, i.e., just keep the friction partner whose surface is tribo-induced negatively charged as the stationary one, can be utilized to suppress the intensity of the tribocharging and the tribodischarging. This work may provide potential applications in numerous areas of science and engineering and also in the everyday life.

---

* Corresponding author. E-mail: xuxuefeng@bjfu.edu.cn

# Ⅰ. Introduction

The tribocharging and the following tribodischarging are common natural phenomena in our daily life, among which include the light induced by the volcanic ash particles [1-2], the glow appeared in an earthquake [3-4], and the spark emitted when a sweater being taken off [5]. The tribocharging and the tribodischarging are generally considered to be harmful. The static electricity can cause damage to the key equipment in the spaceship, which become a serious issue for the space exploration [6-7]. In the powder-handling operations, the electrostatically charged particles can cause fire or even explosion [8]. Some electronic devices such as CMOS-integrated circuits can be destroyed by the electrical discharges induced by the tribocharging [9].

The tribocharging, although harmful in some situations, may be useful in a wide range of applications. In the electrophotography, an image is created after the triboelectrically charged toner particles being attracted to a corona-charged drum [10-11]. Based on the electrostatic properties of the particles that are triboelectrically charged, electrostatic separators can be used for particle separation which can improve the possibilities of recycling wastes [12-13]. By using the triboelectrification and the electrostatic induction, Wang et al. [14-16] have recently invented a series of triboelectric nanogenerators, which can be used to harvest the mechanical energy from our daily life such as walking energy and water wave energy.

Whether we want to utilize or to prevent the tribocharging and the tribodischarging, it is necessary to elucidate the underlying mechanisms. Triboelectrification is the product of the charge transfer between the contacting surfaces. In most cases, it is the electron that moves between the contacting surfaces and results in the tribocharging. For metal-metal contact, the electron will transfer from the surface with a lower work function to the surface with a higher work function [17]. Harper experimentally confirmed that the amount of electron transferred between two different metal spheres is proportional to the difference in the work functions [17-18]. A similar concept, i.e., the effective work function, has been proposed for the insulator, and was then used to explain the electron transfer between the insulator surface and the metal surface [19]. Because electrons can transfer only on the surface rather than inside of the insulators, a surface state model for insulator-insulator contact is introduced, which is used to explain the charging of the particles in the electrophotographic technology [20]. Besides the electron, the tribocharging can also be attributed to the transfer of the ion [21], the proton [22], and even the material [23] between the contacting surfaces.

Once two different material surfaces are brought to rub against each other, the tribocharging often occurs. When discharging, the collision of the accelerated electrons on the air molecules will produce the emission of the light, which is called the triboluminescence (TL), including the ultraviolet (UV) [24-26], the visible [24-30] and the infrared (IR) [24-26]. By comparing the spectrum of the tribo-photons with that of the gas discharge, Nakayama et al. [24-26] proved that the TL is caused by the discharge of the ambient gas in the gap between the contacting surfaces. In addition, another type of TL, the X-ray, generated by the directly impacts of the accelerated electrons on the material surfaces during the tribocharging, was also reported. Strong X-ray has been observed by the peeling of the pressure-sensitive adhesive tape [31], the repeated contact and separation between a silicone rod and a metal-loaded epoxy [32], the continuous friction between a Pb-coated roller and a uPVC tape-coated roller [33], and even in the solid-liquid friction when the meniscus of mercury moving on the glass [34].

Conventional methods such as the grounding method, the isolation method, and the neutralization method are not applicable to suppress the tribocharging and the tribodischarging between insulator surfaces. To explore effective approaches for reducing the tribocharging and the tribodischarging of insulator surfaces, further investigations are needed. However, the tribocharging and the tribodischarging are difficult to be quantitatively measured by the conventional instruments [30]. Because the tribocharging and tribodischarging are often accompanied with the emission of TL, we can investigate the characteristics of the tribocharging and the tribodischarging by measuring the spectrum and the intensity of TL. In the present paper, a disk-on-disk apparatus is first developed for measuring

TL in the sliding friction. The dramatic change in the TL intensity when alter the position of the friction pairs has been observed and then explained by measuring the polarities of the tribo-induced charge on the friction surfaces. Finally, according to the experiment results, a simple and low cost approach which can effectively suppress the tribocharging and the tribodischarging between insulator surfaces has been introduced. The present work may be helpful to the potential application in our daily life and industrial production.

## Ⅱ. Experiments

The sliding friction is carried out by using a disk-on-disk apparatus as showed in Fig.1. In the experiments, the lower disk rotates around its axis and the stationary upper disk is pressed against the rotating disk by a normal force $F_N$. The emitted photons due to the friction are collected by a quartz glass fiber and transferred to a spectrometer iHR320 (HORIBA Scientific). A computer is used to record and analyze the spectrum and the intensity of the light emission. All the experiments in the present paper are conducted under the normal force $F_N = 0.3$ N and the relative sliding velocity V= 0.35 m/s in the ambient air. In the experiments, the room temperature is about 22 ℃ and the relative humidity is about 23% .

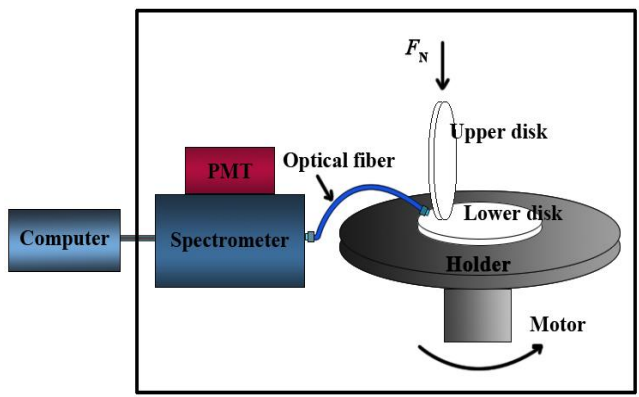

Fig.1 Schematic diagram of the experimental apparatus for measuring TL

The diameter and the thickness of the disks are 30 mm and 2 mm respectively, and the materials of the disks used here are Al$_2$O$_3$ crystal, SiO$_2$ crystal, Al$_2$O$_3$ ceramic and SiO$_2$ glass (Shanghai Daheng

Optical & Fine Mechanics Co. Ltd). Experiments on two separate sets of the friction pairs, one of which consists of $Al_2O_3$ crystal and $SiO_2$ crystal and the other consists of $Al_2O_3$ ceramic and $SiO_2$ glass, were performed. During the experiments, the position of the friction pairs in the same set will be reversed and both the spectrums and the intensity of TL will be measured for each arrangement of the friction pair.

## Ⅲ. Results and Discussion

### 3.1 TL induced by tribodischarging

To clarify the origin of TL during friction, the spectrums of the emitted light in all arrangements of the friction pairs are first measured and then illustrated in Fig. 2. The figure shows that apparent TL is emitted in all the sets of friction pairs except the case of "$SiO_2$ crystal (S) / $Al_2O_3$ crystal (R)", which represents a stationary $SiO_2$ crystal disk on a rotating $Al_2O_3$ crystal disk, here the letter "S" standing for stationary and "R" for rotating. Comparisons between the spectrums reveal that the peaks in all the spectrums, although have different intensities, completely coincides. This implies that the observed TL in all the cases has the same origin.

As suggested by Nakayama [25], the observed TL is generated by the discharging of gas molecules in the gap of the sliding contact, which is in turn induced by the tribocharging on the mating surfaces (see Fig. 3). When friction, an intense electric field in the gap of the contact is formed due to the transfer of the electrons. Under the action of the electric field, the electrons emitted from the negatively charged surface are accelerated and finally collide with the air molecules in the gap. After the collision, when the excited electrons in the molecules fall down to the lower energy level, photons may be emitted. The intensity of TL should be closely related with the intensity of the tribodischarging: the stronger the tribodischarging, the more intensive the TL observed. Thus, we can investigate the intensities of the tribocharging and the tribodischarging by measuring the intensity of the concomitant light.

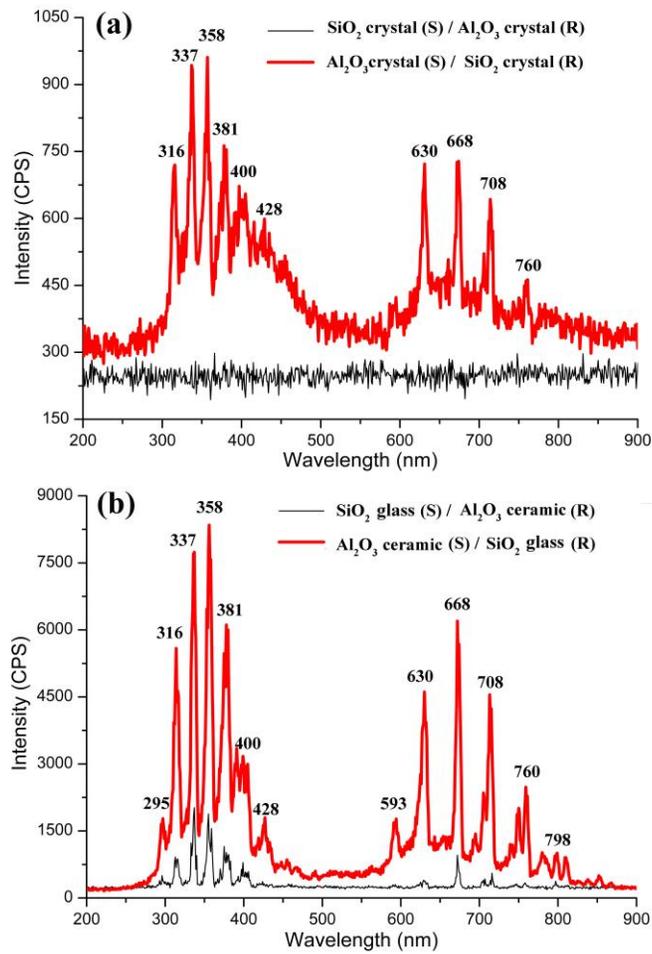

Fig. 2 (a) Spectrums of TL in the friction pair of Al$_2$O$_3$ ceramic and SiO$_2$ glass. (b) Spectrums of TL in the friction pair of Al$_2$O$_3$ crystal and SiO$_2$ crystal.

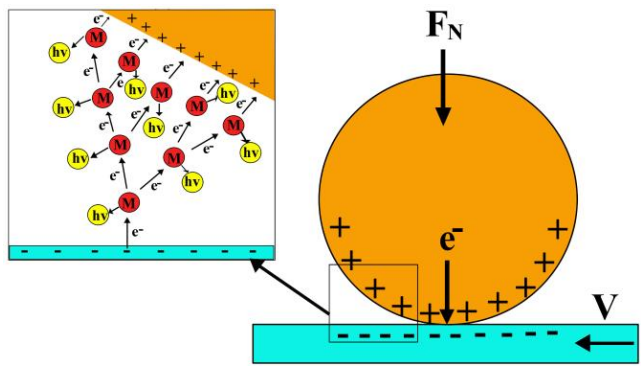

Fig. 3 Mechanism of TL generation.

## 3.2 Effects of reversing the position of the friction pairs

The differences in the peak intensity in the spectrums shown in Fig. 2 imply that changing the spatial position of the friction pair should have influences on the TL, and therefore could have an impact on the tribocharging and the tribodischarging. To observe directly the effects of the spatial configuration of the friction couple, the variation of the TL intensity with time for each arrangement of the friction pair is measured and then depicted in Fig. 4.

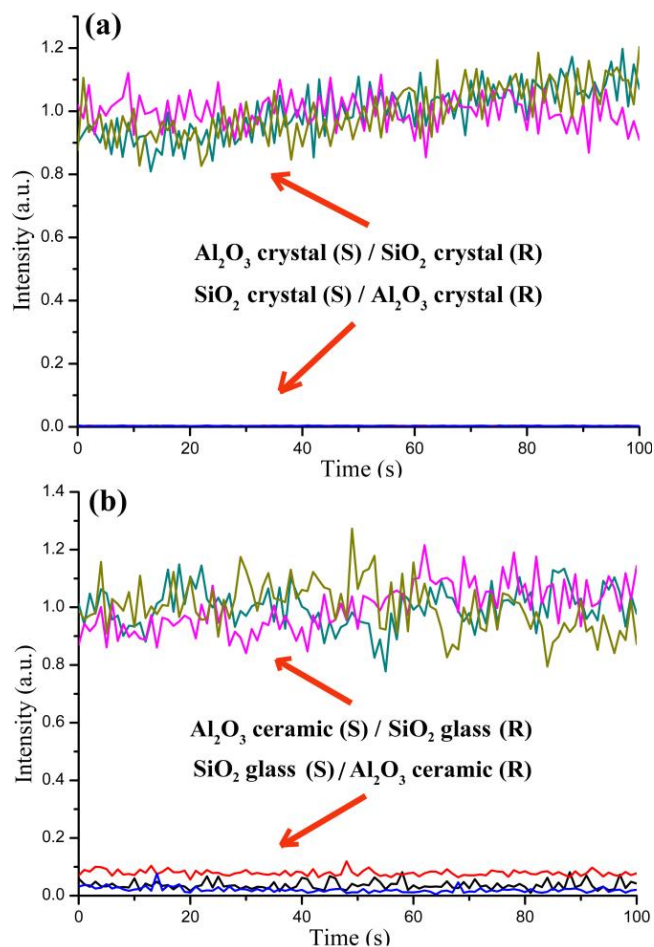

Fig. 4 Variations of TL intensity with time in the friction pair of

(a) $Al_2O_3$ ceramic and fused $SiO_2$ and (b) $Al_2O_3$ crystal and $SiO_2$ crystal.

It can be clearly seen from Fig. 4 that reversing the position of the friction pair does have significant influences on the TL intensity. When a stationary disk of $Al_2O_3$ crystal is pressed on a rotating disk of $SiO_2$ crystal, apparent light is produced. However, there is nearly no light emission if the positions of the two disks are reversed, that is, when a stationary disk of $SiO_2$ crystal is pressed on a rotating disk of Al$_2$O$_3$ crystal (see Fig. 4(b)). For the case of Al$_2$O$_3$ ceramic and SiO$_2$ glass, although TL is observed in both of the spatial configurations of the two disks, the TL intensity is quite different for different spatial configurations. Similar to the friction pairs of two crystal disks, intensive TL emission occurs in the case of a stationary Al$_2$O$_3$ disk and a rotating SiO$_2$ disk, and weak TL is observed when reverse the positions of the disks. The light intensity of the former is at least ten times higher than that of the latter (see Fig. 4(a)). We therefore can conclude that stronger TL and accordingly stronger tribodischarging are generated when the SiO$_2$ disk is rotating and the Al$_2$O$_3$ disk is stationary, and contrarily, weaker TL and weaker tribodischarging are produced when the SiO$_2$ disk is stationary and the Al$_2$O$_3$ disk is rotating.

### 3.3 Reasons for the difference in TL and tribodischarging

As mentioned above, the tribodischarging and consequently the TL are the results of the electric field in the gap of the contact, which is in turn due to the charge transfer (i.e., tribocharging) between the friction surfaces (see Fig. 3). A stronger TL should correspond to a higher electric field, and thus imply more charge transferred. To make clear why the amount of transferred charge will change when just reverse the position of the friction pairs, the friction-induced charging on the mating surfaces has been measured by using Faraday cup method.

Experimental results show that, when Al$_2$O$_3$ disk and SiO$_2$ disk are brought to rub against each other, the Al$_2$O$_3$ disk is always positively charged and the SiO$_2$ disk negatively. This is mainly because the contact potential difference between the two materials. If two objects are put in contact, electrons will transfer from the surface with lower work function to the surface with higher work function, hence the former (the Al$_2$O$_3$ disk here) is positively charged and the latter (the SiO$_2$ disk here) negatively (see Fig. 5(a)).

Besides the contact potential difference, the temperature difference between the two disks at the contact point can also influence the charge transfer. The Seebeck effect indicates that, when two objects at different temperatures contact with each other, the electrons will flow from the hot object to the cool one. Considering the experiments are conducted by using a stationary upper disk pressed on a rotating lower disk, the temperature at the contact point of the upper disk must be always higher than that of the lower disk due to the asymmetric friction. Driven by this temperature gradient, some electrons of the upper disk may transfer to the lower disk (see Fig. 5(b)).

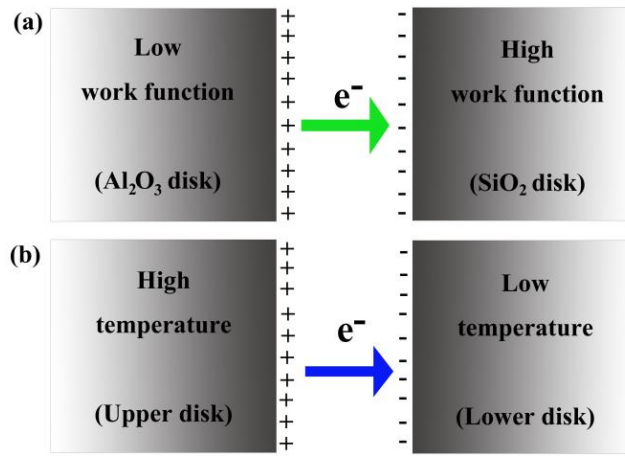

Fig. 5 Direction of the electrons transfer induced by

(a) contact potential difference and (b) Seebeck effect

To explain the intensity difference in the TL and in the tribocharging, it is reasonable to think in terms of the synergy of the contact potential difference and the Seebeck effect on the charge transfer. When the $Al_2O_3$ disk is placed in the above, the charge transfer direction driven by the potential difference (i.e., from $Al_2O_3$ disk to $SiO_2$ disk) and that driven by the temperature gradient (i.e., from the upper disk to the lower disk) are the same (see Fig. 6 (a)). Hence, the quantity of the transferred electrons is large, the strength of the electric field is high, and consequently, the intensity of the tribodischarging and the TL is strong. Contrarily, if the $SiO_2$ disk is placed in the above, the directions of the electron transfer due to the two effects are opposite. This will result in a weak electric field and then a weak tribodischarging and weak TL (see Fig. 6(b)).

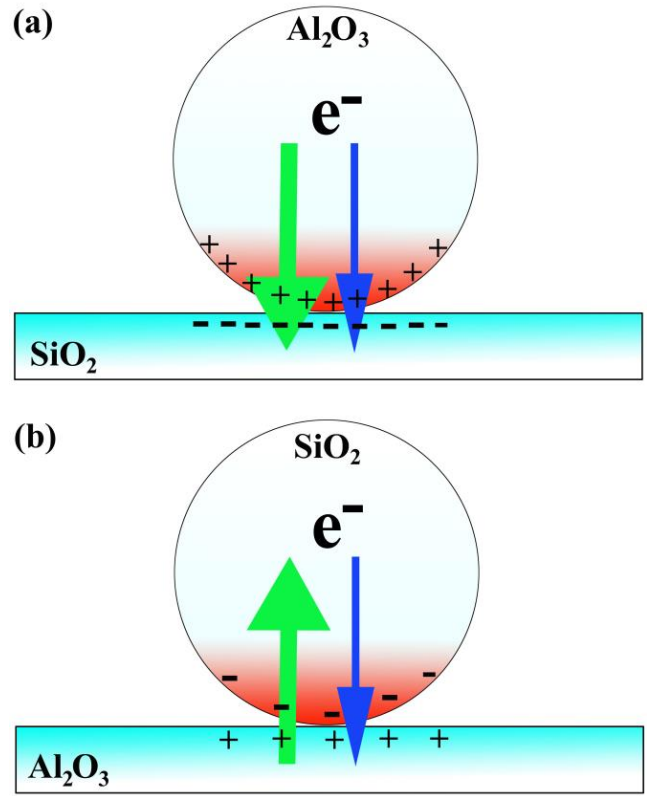

Fig. 6 Combined effects of the contact potential difference and the Seebeck effect on the electron transfer for the cases of (a) strong TL and discharging and (b) weak TL and tribocharging

### 3.4 An approach to modulate the intensity of tribocharging and tribodischarging

Because the tribocharging is the result of the electron transfer between the friction surfaces, changing the quantity of the transferred electrons should alter the intensity of the tribocharging and the TL. From above analyses, a simple approach that can effectively modulate the strength of tribocharging and tribodischarging is proposed here.

To suppress the tribocharging and the tribodischarging, reducing the amount of the transferred charges is needed. For this purpose, the material with higher work function should be chosen as the stationary one of the friction pairs, i.e., the one at a higher temperature due to the asymmetric friction. On the contrary, if we want to enhance the tribocharging and the tribodischarging, the material with lower work function should be chosen as the stationary one of the friction pairs (see Fig. 6).

## Ⅳ. Conclusions

A disk-on-disk apparatus for measuring the spectrum and the intensity of TL in the sliding friction has first been developed. The observed spectrums indicate that the origin of TL here is induced by the discharging of gas molecules in the gap of the sliding contact, which is in turn the result of charge transfer between the mating surfaces. Further experiments show that the TL intensity change significantly when reversing the spatial position of the friction pairs, which implies that it must also have a remarkable effect on the strength of tribocharging and tribodischarging. To clarify the reason for the difference in the TL intensity, the tribo-induced charge on the friction surfaces has further been measured. The results indicate that strong TL, and thus strong tribocharging and tribodischarging, occur when the rotating surface is negatively charged.

By considering the combined effect of the contact potential difference and the Seebeck effect on the electron transfer between the friction surfaces, the apparent difference in the TL intensity has been reasonably explained. Further, an approach that can modulate the intensity of tribocharging and tribodischarging is proposed. We found that simply reversing the relative spatial position of the friction pairs could dramatically alter the strength of tribocharging and tribodischarging. The present work may provide potential applications in numerous fields such as petrochemical industry, power electronics and textile industry.

## Acknowledgements

The work is financially supported by the National Natural Science Foundation of China (Grant NOs 51575054 and 51527901).